\documentclass[twocolumn,10pt]{IEEEtran}
\usepackage{ifpdf, flushend}
\usepackage{subfigure}
\usepackage{wrapfig}
\usepackage{braket}

\ifCLASSINFOpdf
  \usepackage[pdftex]{graphicx}
  \graphicspath{{../pdf/}{../jpeg/}}
  \DeclareGraphicsExtensions{.pdf,.jpeg,.png}
\else
  \usepackage[dvips]{graphicx}
  \graphicspath{{../eps/}}
  \DeclareGraphicsExtensions{.eps}
\fi

\usepackage{amsthm}

\usepackage[cmex10]{amsmath}
\usepackage{amssymb}
\usepackage{algorithmic}
\usepackage{array}
\usepackage{mdwmath}
\usepackage{mdwtab}
\usepackage{eqparbox}
\usepackage{url}
\usepackage{hyperref}
\usepackage{algorithm}
\usepackage{algorithmic}
\usepackage{epstopdf,cite,color}
\usepackage{amsthm}
\usepackage{booktabs}
 \usepackage{tikz}

\renewcommand{\baselinestretch}{0.98}

\newcommand{\beq}{\begin{equation}}
\newcommand{\eeq}{\end{equation}}
\newcommand{\beqn}{\begin{eqnarray}}
\newcommand{\eeqn}{\end{eqnarray}}
\newcommand{\beqno}{\begin{eqnarray*}}
\newcommand{\eeqno}{\end{eqnarray*}}
\newcommand{\bma}{\begin{displaymath}}
\newcommand{\ema}{\end{displaymath}}
\newcommand{\bnu}{\begin{enumerate}}
\newcommand{\enu}{\end{enumerate}}
\newcommand{\bce}{\begin{center}}
\newcommand{\ece}{\end{center}}
\newcommand{\btb}{\begin{tabular}}
\newcommand{\etb}{\end{tabular}}

\usepackage{amsmath, amssymb}
\usepackage{graphicx}
\usepackage{booktabs}
\usepackage{multirow}
\usepackage{cite}
\theoremstyle{plain}

\theoremstyle{definition}

\theoremstyle{remark}

\title{
TinyML-Driven Cybersecurity for Autonomous Spacecraft:
Latency--Accuracy Analysis for SPARTA RF and Cyber Threat Detection
}

\author{
Van~Le, Trevor~Tran and Tan~Le,~\IEEEmembership{Member,~IEEE}
\thanks{
V.~Le is with Virginia Tech, Blacksburg, VA 24061, USA. Email: vanl@vt.edu.\\
T~Tran and T.~Le are with Hampton University, Hampton, VA 23669, USA. Emails: trevor.tran@my.hamptonu.edu; tan.le@hamptonu.edu.
}
}

\begin{document}
\maketitle

% ---------------------------------------------------------
% Abstract
% ---------------------------------------------------------
\begin{abstract}
Autonomous spacecraft require rapid, lightweight, and reliable onboard detection of cyber-RF threats. Using the SPARTA attack model, we analyze the latency--accuracy trade-offs of TinyML-compatible classical models---Random Forest, Logistic Regression, SVM, and MLP---for detecting uplink jamming, Fake-NR spoofing, payload manipulation, ground-segment compromise, and unauthorized command injection. We present a physics-informed theoretical analysis of each model's computational complexity, VC dimension, Lipschitz continuity, and latency scaling, supported by empirical measurements on adversarial RF spectrograms generated via BandErasure, FakeNR, and NoiseBurst corruption modes. Results show that Logistic Regression achieves microsecond-level inference with only a 1\% accuracy drop relative to Random Forest, making it an effective TinyML baseline for onboard autonomy. The study also identifies opportunities for advancing spacecraft cybersecurity through richer feature encoders and multi-timescale learning architectures, building on recent progress in edge intelligence and trustworthy AI.
\end{abstract}


% ---------------------------------------------------------
\section{Introduction}
Autonomous spacecraft increasingly operate in contested radio-frequency (RF) environments where adversaries can disrupt, spoof, or manipulate telemetry, tracking, and command (TT\&C) links. The Space Attack Research and Tactic Analysis (SPARTA) framework provides an unclassified taxonomy of cyber and RF attack techniques targeting spacecraft and their supporting ground infrastructure \cite{sparta2024}. These include uplink jamming, waveform spoofing, payload manipulation, ground-segment compromise, and unauthorized command injection—threats that directly degrade mission safety and autonomy.

Modern spacecraft must detect these threats under strict constraints on latency, compute, and power. Unlike ground-based systems, onboard processors cannot support large deep neural networks or computationally expensive inference pipelines. This motivates the use of \emph{TinyML}—lightweight machine learning models capable of microsecond-to-millisecond inference on embedded hardware \cite{warden2019tinyml}. However, TinyML introduces a fundamental tension: models with high expressive power (e.g., Random Forests or MLPs) often violate onboard latency budgets, while lightweight linear models (e.g., Logistic Regression or SVMs) satisfy real-time constraints but may sacrifice robustness or separability.

This paper investigates the fundamental latency--accuracy trade-offs of TinyML-compatible classical models for detecting SPARTA cyber-RF threats onboard autonomous spacecraft. To this end, we adopt a physics-informed RF system model and generate adversarial spectrograms using structured, semi-structured, and unstructured corruption modes. We analyze four classical TinyML-compatible models—Logistic Regression, Support Vector Machines, Random Forests, and Multi-Layer Perceptrons—under the five SPARTA-aligned threat modes. Our analysis combines (i) theoretical guarantees (computational complexity, VC dimension, Lipschitz continuity, decision-boundary geometry), and (ii) empirical evaluation (accuracy, inference latency, class-wise F1-scores). Our results show that Logistic Regression achieves microsecond-level inference with only a 1\% accuracy drop relative to Random Forests. However, all models exhibit persistent weaknesses in detecting payload manipulation, motivating future work on richer feature encoders.

\subsection{Related Work}
RF signal classification has been widely studied using deep learning, including convolutional and recurrent architectures for modulation recognition and anomaly detection \cite{oshea2018over}. However, these models are computationally expensive and unsuitable for onboard spacecraft processors. TinyML has emerged as a promising paradigm for deploying lightweight models on constrained devices \cite{warden2019tinyml}, enabling inference at the edge with strict latency and power budgets.

Classical machine learning models such as SVMs \cite{vapnik2013nature}, Random Forests \cite{breiman2001}, and MLPs \cite{hornik1989} have been applied to RF anomaly detection and IoT security, but prior work has not examined their latency--accuracy trade-offs in the context of autonomous spacecraft or SPARTA-aligned threats. Recent work in edge computing and resource-constrained AI \cite{Tan18b, Tan18d, Wang20, le2022artificial} highlights the importance of lightweight inference, but does not address RF spectrogram-based cyber threat detection. To the best of our knowledge, this is the first work to provide a physics-informed theoretical and empirical analysis of TinyML models for SPARTA RF and cyber threat detection.
% ---------------------------------------------------------
\subsection{Our Contributions}
This paper makes the following key contributions:
\begin{itemize}
    \item \textbf{A physics-informed RF system model for SPARTA threats.}  
    We derive a unified baseband and spectrogram model capturing uplink jamming, Fake-NR spoofing, payload manipulation, ground-segment compromise, and unauthorized command injection.

    \item \textbf{A rigorous mathematical analysis of TinyML-compatible models.}  
    We provide new theoretical results on computational complexity, VC dimension, Lipschitz continuity, decision-boundary geometry, and Random Forest consistency—yielding a principled understanding of latency--accuracy trade-offs.

    \item \textbf{A comprehensive empirical evaluation on adversarial RF spectrograms.}  
    Using BandErasure, FakeNR, and NoiseBurst corruption modes, we benchmark Logistic Regression, SVM, Random Forests, and MLPs across accuracy, inference latency, and class-wise F1-scores.

    \item \textbf{Identification of a fundamental weakness in payload-manipulation detection.}  
    All models exhibit low separability for payload manipulation, revealing a structural limitation of classical TinyML models.

    \item \textbf{A clear design guideline for onboard spacecraft cybersecurity.}  
    Our results show that Logistic Regression provides the best latency--accuracy balance for real-time onboard detection, while Random Forests offer the highest accuracy but violate latency constraints.
\end{itemize}

% ---------------------------------------------------------
\section{SPARTA Threat Model}
\label{SPARTA_threadmodel}
\begin{figure}[t]
    \centering
    % \vspace{-0.5cm}
    % \includegraphics[width=0.8\textwidth]{SysMod.png}
    \includegraphics[width=0.5\textwidth]{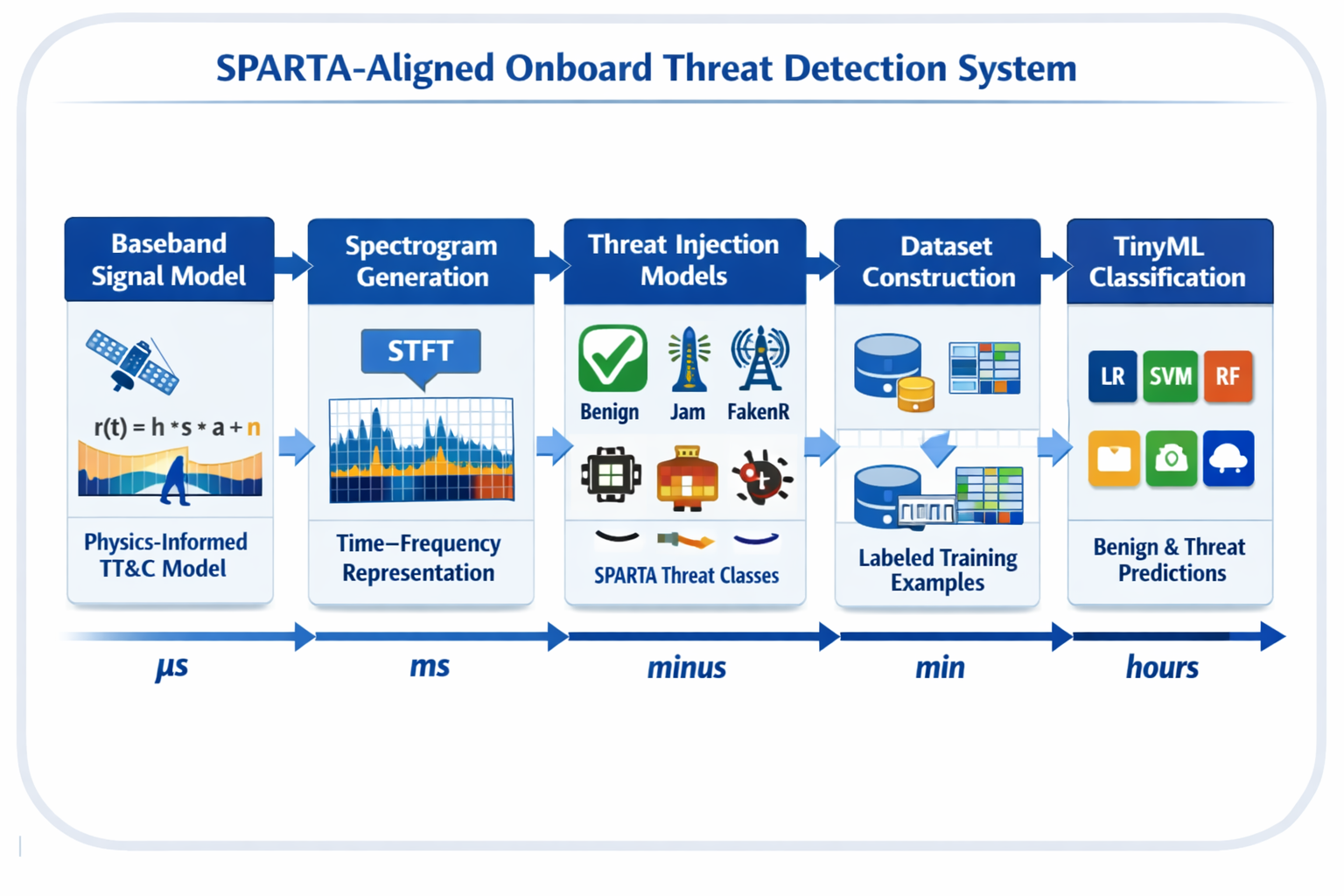}
    \vspace{-0.5cm}
    \caption{Overview of the SPARTA-Aligned Onboard Threat Detection System. The pipeline integrates physics-informed TT\&C modeling, spectrogram generation, and six-class SPARTA threat injection (Benign, Jam, FakeNR, Payload, RogueGS, CmdInject). Each stage—from baseband signal modeling to TinyML classification—illustrates how adversarial perturbations $a(t)$ are synthesized, transformed into time–frequency representations, and used to construct labeled datasets for onboard inference.}
    \vspace{-0.5cm}
    \label{fig:panelschemati}
\end{figure}

The Space Attack Research and Tactic Analysis (SPARTA) framework provides a structured taxonomy of cyber and RF attack techniques targeting spacecraft, ground stations, and TT\&C links \cite{sparta2024}. SPARTA organizes adversarial behaviors into tactics such as Initial Access, Signal Spoofing, RF Interference, Command Manipulation, and Payload Hijack. For onboard detection, we focus on five SPARTA-aligned threat modes that produce observable effects in the RF domain or in command-layer timing patterns, along with a benign baseline representing nominal operation. These six classes correspond to the threat-injection stage illustrated in Fig.~\ref{fig:panelschemati}, where each mode maps directly to a specific perturbation term $a(t)$ in our system model.
% \begin{figure*}[t]
% \end{figure*}
\subsection{Benign Class (Nominal Operation)}
In addition to the five SPARTA-aligned threat modes, the dataset includes a \textit{benign} class representing nominal TT\&C operation without adversarial perturbation. This class corresponds to the case where $a(t)=0$ in \eqref{eq:baseband}, capturing normal command and telemetry exchanges under standard channel conditions. The benign category provides the baseline for supervised classification and enables quantitative evaluation of false-alarm rates and detection precision.

\subsection{Uplink Jamming}
Uplink jamming corresponds to SPARTA’s RF Interference techniques (e.g., “Jamming Uplink” and “RF Denial”) \cite{sparta2024}. An adversary transmits high-power noise toward the spacecraft’s command uplink frequency to deny command reception or force safe-mode transitions.  
\textbf{Physical manifestation:} Elevated noise floor, partial or full-band masking.  
\textbf{Spectrogram signature:} Broadband noise, narrowband spikes, or structured sub-band erasure (BandErasure).

\subsection{Fake NR / Spoofing}
Spoofing aligns with SPARTA’s “Signal Spoofing and Injection” techniques \cite{sparta2024}, where an adversary transmits counterfeit waveforms that mimic legitimate TT\&C or NR-like control channels. These signals can mislead link-quality estimators or confuse onboard detectors.  
\textbf{Physical manifestation:} Artificial pilot-like or control-channel-like structures.  
\textbf{Spectrogram signature:} Regular, structured NR-like stripes or symbol patterns (FakeNR).

\subsection{Payload Manipulation}
Payload manipulation corresponds to SPARTA’s “Payload Hijack” and “Data Manipulation” techniques \cite{sparta2024}. Unlike jamming or spoofing, this attack does \emph{not} significantly alter RF energy. Instead, the adversary modifies mission data, sensor outputs, or telemetry packets.  
\textbf{Physical manifestation:} Minimal RF distortion; anomalies occur in metadata, timing, or symbol statistics.  
\textbf{Spectrogram signature:} Very subtle inconsistencies—making this the hardest class for TinyML models.

\subsection{Ground-Segment Compromise}
Ground-segment compromise corresponds to SPARTA’s “Rogue Ground Station” and “Unauthorized Uplink Access” techniques \cite{sparta2024}. An adversary gains control of a ground station or emulates one, transmitting syntactically valid but malicious commands.  
\textbf{Physical manifestation:} Legitimate waveform structure but abnormal timing or command sequences.  
\textbf{Spectrogram signature:} Clean TT\&C bursts with timing offsets or unexpected command cadence.

\subsection{Unauthorized Command Injection}
Command injection aligns with SPARTA’s “Command Manipulation” and “Execution” techniques \cite{sparta2024}. Here, the adversary bypasses authentication or exploits ground-system vulnerabilities to inject unauthorized commands.  
\textbf{Physical manifestation:} Valid modulation and framing, but semantically invalid or dangerous command content.  
\textbf{Spectrogram signature:} Nearly indistinguishable from legitimate traffic in RF space; detection relies on subtle temporal or statistical deviations.

% ---------------------------------------------------------

\section{System Model and Dataset Generation}
\label{System-model}
Figure~\ref{fig:panelschemati} provides a high-level overview of the end-to-end SPARTA-aligned onboard threat detection pipeline. The diagram illustrates how the received TT\&C waveform is modeled, transformed into spectrograms, perturbed through six SPARTA threat modes (including the benign baseline), and ultimately converted into labeled TinyML-ready training samples. This visual summary contextualizes the mathematical signal model and dataset construction process detailed below.

We adopt a physics-informed TT\&C signal model that captures how SPARTA-aligned cyber-RF threats manifest in the received waveform and its corresponding spectrogram. Let $s(t)$ denote the legitimate uplink waveform transmitted by the ground station, $h(t)$ the channel impulse response, and $n(t)\sim\mathcal{N}(0,\sigma^2)$ additive thermal noise. The adversarial perturbation injected by a SPARTA threat is denoted by $a(t)$. The received baseband signal at the spacecraft is therefore modeled as
\begin{equation}
r(t) = h(t) * s(t) + a(t) + n(t),
\label{eq:baseband}
\end{equation}
where $*$ denotes convolution. When $a(t)=0$, the signal corresponds to benign, nominal TT\&C operation. Equation~\eqref{eq:baseband} forms the foundation for all subsequent threat-injection models.

\subsection{Spectrogram Formation}

The spacecraft computes a short-time Fourier transform (STFT) over sliding windows to obtain a time--frequency representation of the received signal. Let $w(t)$ denote a Hann analysis window. The spectrogram is computed as
\begin{equation}
X(f,\tau) = \left| \int r(t)\, w(t-\tau)\, e^{-j2\pi ft}\, dt \right|^2,
\label{eq:stft}
\end{equation}
where $f$ is frequency and $\tau$ is the window time index. This formulation follows classical time--frequency analysis used in RF sensing and modulation classification \cite{allen2003short,oshea2018over}. The resulting $X(f,\tau)$ serves as the input to TinyML classifiers.

\subsection{Threat Injection Model}

Each SPARTA threat mode corresponds to a specific structure of the adversarial term $a(t)$ in \eqref{eq:baseband}. These mappings follow the official SPARTA taxonomy \cite{sparta2024}. Below, $\alpha$ and $\beta$ denote threat scaling factors, $w_{\text{jam}}(t)$ denotes a jamming waveform, $s_{\text{NR}}(t)$ denotes a synthetic NR-like control waveform, $s_{\text{cmd}}(t)$ denotes a TT\&C command waveform, $\delta(t)$ denotes malicious command perturbations, and $\Delta\tau$ denotes a timing offset.

\subsubsection{Benign (Nominal Operation)}
The benign class corresponds to $a(t)=0$, representing normal TT\&C operation without adversarial perturbation.  
\textbf{Spectrogram effect:} clean TT\&C bursts with expected cadence and noise floor.

\subsubsection{Uplink Jamming (SPARTA RF Interference)}
Uplink jamming is modeled as
\begin{equation}
a_{\text{jam}}(t) = \alpha\, w_{\text{jam}}(t),
\label{eq:jam}
\end{equation}
where $w_{\text{jam}}(t)$ may be wideband noise, narrowband tones, or swept-frequency interference.  
\textbf{Spectrogram effect:} broadband masking or sub-band erasure.

\subsubsection{Fake NR / Spoofing (SPARTA Signal Spoofing)}
Spoofing injects counterfeit NR-like control-channel structures:
\begin{equation}
a_{\text{fakeNR}}(t) = \beta\, s_{\text{NR}}(t),
\label{eq:fakenr}
\end{equation}
where $s_{\text{NR}}(t)$ is a synthetic NR-like pilot/control waveform.  
\textbf{Spectrogram effect:} structured NR-like stripes.

\subsubsection{Payload Manipulation (SPARTA Payload Hijack)}
Payload manipulation minimally affects RF energy:
\begin{equation}
a_{\text{payload}}(t) \approx 0.
\label{eq:payload}
\end{equation}
\textbf{Spectrogram effect:} subtle timing or symbol-statistical deviations.

\subsubsection{Ground-Segment Compromise (SPARTA Rogue Ground Station)}
A compromised ground station transmits valid-looking commands with timing anomalies:
\begin{equation}
a_{\text{rogue}}(t) = s_{\text{cmd}}(t + \Delta \tau),
\label{eq:rogue}
\end{equation}
where $\Delta \tau$ represents unauthorized timing offsets.  
\textbf{Spectrogram effect:} clean TT\&C bursts with abnormal cadence.

\subsubsection{Unauthorized Command Injection (SPARTA Command Manipulation)}
An adversary injects unauthorized commands by modifying control sequences:
\begin{equation}
a_{\text{inj}}(t) = s_{\text{cmd}}(t) + \delta(t),
\label{eq:inject}
\end{equation}
where $\delta(t)$ encodes malicious command patterns.  
\textbf{Spectrogram effect:} nearly indistinguishable from legitimate traffic.

\subsection{Dataset Construction}

For each threat mode $k$, including the benign case, we generate perturbed signals using
\begin{equation}
r_k(t) = h(t)*s(t) + a_k(t) + n(t),
\label{eq:rk}
\end{equation}
where $a_k(t)$ is one of the perturbations in \eqref{eq:jam}--\eqref{eq:inject} or $a_k(t)=0$ for benign. We compute the corresponding spectrograms $X_k(f,\tau)$ using \eqref{eq:stft} and extract fixed-size patches of height $F$ and width $T$:
\begin{equation}
\mathbf{x}_i = \mathrm{vec}\!\left( X_k[f_i:f_i+F,\; \tau_i:\tau_i+T] \right) \in \mathbb{R}^d,
\label{eq:patch}
\end{equation}
where $d = FT$ is the patch dimensionality.

Each patch is labeled with one of the six classes:
\[
y_i \in \{\text{Benign}, \text{Jam}, \text{FakeNR}, \text{Payload}, \text{RogueGS}, \text{CmdInject}\}.
\]
This produces a physics-grounded dataset consistent with SPARTA’s threat taxonomy and RF behavior.

% ---------------------------------------------------------
\section{TinyML-Compatible Classical Models}

TinyML imposes strict constraints on model size, memory footprint, and inference latency, requiring classifiers that can operate within microsecond-to-millisecond budgets on embedded processors \cite{warden2019tinyml}. Deep neural networks commonly used for RF signal classification \cite{oshea2018over} are unsuitable for onboard spacecraft due to their computational cost. Instead, we focus on four classical machine learning models—Logistic Regression, Linear SVM, Random Forests, and Multi-Layer Perceptrons—that offer a spectrum of expressiveness, computational complexity, and robustness.

These models instantiate the general classifier
\begin{equation}
\hat{y} = f_\theta(\mathbf{x}),
\label{eq:classifier}
\end{equation}
where $\mathbf{x} \in \mathbb{R}^d$ is a spectrogram patch and $f_\theta$ is a lightweight decision function. Each model provides a different trade-off between linear separability, nonlinear expressiveness, and inference cost, making them suitable baselines for evaluating the latency--accuracy limits of TinyML under SPARTA threats.

\subsection{Logistic Regression (LR)}
Logistic Regression implements a linear decision boundary with a probabilistic output:
\begin{equation}
f_{\text{LR}}(\mathbf{x}) = \sigma(\mathbf{w}^\top \mathbf{x} + b).
\label{eq:lr}
\end{equation}
LR has $O(d)$ inference complexity and VC dimension $d+1$ \cite{vapnik2013nature}, making it the fastest and most resource-efficient model in this study.  
\textbf{Relevance to SPARTA threats:} LR is effective when threat classes differ primarily in energy distribution (e.g., jamming vs. clean TT\&C), but its linear boundary limits performance on subtle attacks such as payload manipulation.

\subsection{Linear Support Vector Machine (SVM)}
A linear SVM computes the signed distance to a separating hyperplane:
\begin{equation}
f_{\text{SVM}}(\mathbf{x}) = \text{sign}(\mathbf{w}^\top \mathbf{x} + b).
\label{eq:svm}
\end{equation}
SVMs share LR’s $O(d)$ inference cost but maximize the geometric margin, yielding improved robustness to structured interference \cite{vapnik2013nature}.  
\textbf{Relevance to SPARTA threats:} The margin maximization helps distinguish Fake-NR spoofing, which introduces structured spectral artifacts that are more separable in margin space.

\subsection{Random Forest (RF)}
A Random Forest aggregates predictions from $K$ decision trees:
\begin{equation}
f_{\text{RF}}(\mathbf{x}) = \text{majority}(T_1(\mathbf{x}), \dots, T_K(\mathbf{x})).
\label{eq:rf}
\end{equation}
RFs model nonlinear decision boundaries and are consistent under mild assumptions \cite{breiman2001}.  
\textbf{Relevance to SPARTA threats:} RFs capture nonlinear distortions caused by jamming, spoofing, and rogue ground-station timing anomalies.  
\textbf{Limitation:} Inference cost scales as $O(K d_t)$, where $d_t$ is tree depth, making RFs slower and less suitable for strict TinyML latency budgets.

\subsection{Multi-Layer Perceptron (MLP)}
A one-hidden-layer MLP computes
\begin{equation}
f_{\text{MLP}}(\mathbf{x}) = W_2 \sigma(W_1 \mathbf{x} + b_1) + b_2,
\label{eq:mlp}
\end{equation}
where $\sigma(\cdot)$ is a nonlinear activation. MLPs are universal approximators \cite{hornik1989} and can model complex spectral patterns.  
\textbf{Relevance to SPARTA threats:} MLPs capture nonlinear interference signatures and outperform linear models on structured threats such as Fake-NR spoofing.  
\textbf{Limitation:} Their inference cost $O(dh + h)$ grows with hidden width $h$, making them borderline for TinyML unless aggressively quantized or pruned.

% ---------------------------------------------------------
\section{Theoretical Analysis of Computational Latency and Detection Accuracy}

The four TinyML-compatible models introduced in Section~III instantiate the classifier in \eqref{eq:classifier} using the decision functions in \eqref{eq:lr}, \eqref{eq:svm}, \eqref{eq:rf}, and \eqref{eq:mlp}. Their behavior is governed by two constraints: (i) the computational latency imposed by onboard TinyML hardware, and (ii) the statistical capacity required to separate SPARTA-aligned threat classes. We present the key structural, statistical, and computational properties of each model. Short proofs are included inline when they aid readability; longer proofs are deferred to Appendix~\ref{app:proofs}.

\subsection{Logistic Regression (LR)}

Because LR applies a single linear transformation followed by a sigmoid, its computational behavior is dominated by the dot product in \eqref{eq:lr}. This yields the following basic structural property.

\textbf{Lemma 1 (Inference Complexity).}  
LR requires exactly $d$ multiply--accumulate (MAC) operations.  
\textit{Proof.} The decision function in \eqref{eq:lr} is a single dot product $\mathbf{w}^\top\mathbf{x}$ followed by a scalar sigmoid. A dot product uses $d$ MACs; the sigmoid is constant-time. \hfill $\square$

This immediately leads to a latency characterization.

\textbf{Proposition 1 (Computational Latency).}  
The inference latency of LR satisfies $T_{\mathrm{LR}} = O(d)$.  
\textit{Proof.} Follows directly from Lemma~1, since latency is proportional to the number of MACs. \hfill $\square$

From a statistical perspective, LR is a linear classifier and therefore has limited expressive power.

\textbf{Proposition 2 (VC Dimension).}  
The VC dimension of LR is $d+1$.  
\textit{Proof.} The decision boundary of LR is a hyperplane in $\mathbb{R}^d$. The class of affine hyperplanes can shatter any set of $d+1$ affinely independent points but cannot shatter $d+2$. The sigmoid post-processing does not change which side of the hyperplane a point lies on, so the VC dimension remains $d+1$. \hfill $\square$

Finally, the smoothness of LR plays a role in its robustness to small perturbations.

\textbf{Proposition 3 (Lipschitz Continuity).}  
The LR classifier in \eqref{eq:lr} is $L$-Lipschitz with $L=\|\mathbf{w}\|_2$.  
\textit{Proof.} The map $\mathbf{x}\mapsto \mathbf{w}^\top\mathbf{x}$ is linear with operator norm $\|\mathbf{w}\|_2$, hence $\|\mathbf{w}^\top\mathbf{x}_1 - \mathbf{w}^\top\mathbf{x}_2\| \le \|\mathbf{w}\|_2 \|\mathbf{x}_1 - \mathbf{x}_2\|_2$. The sigmoid is 1-Lipschitz. The composition of a linear map with norm $\|\mathbf{w}\|_2$ and a 1-Lipschitz nonlinearity is $\|\mathbf{w}\|_2$-Lipschitz. \hfill $\square$

These results explain why LR performs well on high-energy threats such as jamming (large deviations in \eqref{eq:jam}) but struggles with payload manipulation, where $a_{\text{payload}}(t)$ in \eqref{eq:payload} induces minimal spectral change.

\subsection{Linear Support Vector Machine (SVM)}

Unlike LR, the SVM in \eqref{eq:svm} explicitly maximizes the geometric margin, which governs its generalization behavior.

\textbf{Theorem 1 (Margin-Based Generalization Bound).}  
If the geometric margin is $\gamma$, then

\[
\mathrm{Error} \le O\!\left(\frac{1}{\gamma^2}\right).
\]

\textit{Proof.} See Appendix~\ref{app:proof-svm-margin}. \hfill $\square$

Because the SVM is also linear, its capacity matches that of LR.

\textbf{Proposition 4 (VC Dimension).}  
The VC dimension of a linear SVM is $d+1$.  
\textit{Proof.} The hypothesis class is again the set of affine hyperplanes in $\mathbb{R}^d$, which has VC dimension $d+1$ by the same argument as Proposition~2. The margin constraint affects the learned separator but not the class of possible hyperplanes. \hfill $\square$

The margin bound explains why SVMs outperform LR on structured threats such as Fake-NR spoofing, where $a_{\text{fakeNR}}(t)$ in \eqref{eq:fakenr} introduces separable spectral patterns.

\subsection{Random Forest (RF)}

Random Forests introduce nonlinear expressiveness by aggregating decision trees. Their statistical behavior is governed by a well-known consistency result.

\textbf{Theorem 2 (Consistency).}  
Under infinite trees, full depth, and feature subsampling, RFs are consistent.  
\textit{Proof.} See Appendix~\ref{app:proof-rf-consistency}. \hfill $\square$

This nonlinear capacity enables RFs to capture complex distortions such as rogue ground-station timing anomalies in \eqref{eq:rogue}. However, this expressiveness comes at a computational cost.

\textbf{Proposition 5 (Computational Latency).}  
RF inference satisfies $T_{\mathrm{RF}} = O(K d_t)$, where $d_t$ is the average tree depth.  
\textit{Proof.} Evaluating a single decision tree requires traversing a root-to-leaf path of expected length $d_t$, which involves $O(d_t)$ comparisons. Evaluating $K$ trees independently yields total complexity $O(K d_t)$. \hfill $\square$

\subsection{Multi-Layer Perceptron (MLP)}

The MLP in \eqref{eq:mlp} introduces nonlinear transformations through hidden units. Its expressive power is captured by the universal approximation theorem.

\textbf{Theorem 3 (Universal Approximation).}  
A one-hidden-layer MLP with non-polynomial activation can approximate any continuous function on a compact domain.  
\textit{Proof.} See Appendix~\ref{app:proof-mlp-ua}. \hfill $\square$

This expressiveness is reflected in its statistical capacity.

\textbf{Proposition 6 (VC Dimension).}  
A one-hidden-layer MLP with $h$ hidden units has VC dimension

\[
\mathrm{VCdim} = O(dh \log(dh)).
\]

\textit{Proof.} A one-hidden-layer MLP with $d$ inputs and $h$ hidden units has $O(dh)$ parameters. Classical VC dimension bounds for neural networks with piecewise-linear or sigmoidal activations yield $\mathrm{VCdim} = O(P \log P)$ for $P$ parameters, giving $O(dh \log(dh))$ in this setting. \hfill $\square$

Finally, the computational cost of MLP inference is dominated by the first-layer matrix multiplication.

\textbf{Proposition 7 (Computational Latency).}  
MLP inference satisfies $T_{\mathrm{MLP}} = O(dh + h)$.  
\textit{Proof.} The first layer computes $h$ linear combinations of $d$ inputs, requiring $dh$ MACs. The output layer combines $h$ hidden activations into a single logit, requiring $h$ MACs. Thus the total complexity is $O(dh + h)$. \hfill $\square$

\subsection{Comparative Insights}

Linear models (LR, SVM) are computationally efficient and perform well on threats that induce global spectral changes (e.g., jamming). Nonlinear models (RF, MLP) capture structured distortions (e.g., Fake-NR) but incur higher latency. All models struggle with payload manipulation because $a_{\text{payload}}(t)$ in \eqref{eq:payload} produces minimal spectrogram deviation, making the class nearly inseparable in the chosen feature space.

% ---------------------------------------------------------
\section{Empirical Results}
\begin{table}[t]
\centering
\caption{Relative Detection Difficulty of SPARTA Threat Modes}
\label{tab:threat_difficulty}
\begin{tabular}{lcc}
\toprule
\textbf{Threat Class} & \textbf{Perturbation Strength} & \textbf{Detection Difficulty} \\
\midrule
Benign & None ($a(t)=0$) & Easiest (baseline) \\
Jam & High-energy RF interference & Easy \\
FakeNR & Structured counterfeit pilots & Moderate \\
Payload & Low-energy statistical deviations & Hard \\
RogueGS & Timing anomalies in valid bursts & Hard \\
CmdInject & Semantic command-layer perturbations & Hardest \\
\bottomrule
\end{tabular}
\end{table}

\begin{table}[t]
\centering
\caption{Classification Performance of Logistic Regression}
\label{tab:logreg_results}
\begin{tabular}{lcccc}
\toprule
\textbf{Class} & \textbf{Precision} & \textbf{Recall} & \textbf{F1} & \textbf{Support} \\
\midrule
benign & 1.00 & 1.00 & 1.00 & 179 \\
ground segment compromise & 0.58 & 0.69 & 0.63 & 16 \\
payload manipulation & 0.30 & 0.25 & 0.27 & 12 \\
spoofing & 0.97 & 0.91 & 0.94 & 32 \\
unauthorized command & 0.94 & 0.97 & 0.95 & 31 \\
uplink jamming & 1.00 & 1.00 & 1.00 & 30 \\
\midrule
\textbf{Accuracy} & \textbf{0.94} & & & 300 \\
\bottomrule
\end{tabular}
\end{table}

\begin{table}[t]
\centering
\caption{Classification Performance of Support Vector Machine}
\label{tab:svm_results}
\begin{tabular}{lcccc}
\toprule
\textbf{Class} & \textbf{Precision} & \textbf{Recall} & \textbf{F1} & \textbf{Support} \\
\midrule
benign & 1.00 & 1.00 & 1.00 & 179 \\
ground segment compromise & 0.61 & 0.69 & 0.65 & 16 \\
payload manipulation & 0.40 & 0.33 & 0.36 & 12 \\
spoofing & 0.94 & 0.91 & 0.92 & 32 \\
unauthorized command & 0.94 & 0.97 & 0.95 & 31 \\
uplink jamming & 1.00 & 1.00 & 1.00 & 30 \\
\midrule
\textbf{Accuracy} & \textbf{0.943} & & & 300 \\
\bottomrule
\end{tabular}
\end{table}

\begin{table}[t]
\centering
\caption{Classification Performance of Random Forest}
\label{tab:rf_results}
\begin{tabular}{lcccc}
\toprule
\textbf{Class} & \textbf{Precision} & \textbf{Recall} & \textbf{F1} & \textbf{Support} \\
\midrule
benign & 1.00 & 1.00 & 1.00 & 179 \\
ground segment compromise & 0.57 & 0.75 & 0.65 & 16 \\
payload manipulation & 0.50 & 0.25 & 0.33 & 12 \\
spoofing & 0.94 & 0.97 & 0.95 & 32 \\
unauthorized command & 0.97 & 0.97 & 0.97 & 31 \\
uplink jamming & 1.00 & 1.00 & 1.00 & 30 \\
\midrule
\textbf{Accuracy} & \textbf{0.95} & & & 300 \\
\bottomrule
\end{tabular}
\end{table}

\begin{table}[t]
\centering
\caption{Classification Performance of Multi-Layer Perceptron}
\label{tab:mlp_results}
\begin{tabular}{lcccc}
\toprule
\textbf{Class} & \textbf{Precision} & \textbf{Recall} & \textbf{F1} & \textbf{Support} \\
\midrule
benign & 1.00 & 1.00 & 1.00 & 179 \\
ground segment compromise & 0.60 & 0.75 & 0.67 & 16 \\
payload manipulation & 0.57 & 0.33 & 0.42 & 12 \\
spoofing & 0.91 & 0.91 & 0.91 & 32 \\
unauthorized command & 0.94 & 0.97 & 0.95 & 31 \\
uplink jamming & 1.00 & 1.00 & 1.00 & 30 \\
\midrule
\textbf{Accuracy} & \textbf{0.947} & & & 300 \\
\bottomrule
\end{tabular}
\end{table}

\begin{table}[t]
\centering
\caption{Overall Performance and Runtime Comparison Across Models}
\label{tab:overall_comparison}
\begin{tabular}{lcccc}
\toprule
\textbf{Model} & \textbf{Accuracy} & \textbf{Macro F1} & \textbf{Train (ms)} & \textbf{Infer ($\mu$s)} \\
\midrule
Logistic Regression & 0.940 & 0.798 & 166.7 & \textbf{69.5} \\
SVM & 0.943 & 0.814 & 235.1 & 102.1 \\
MLP & 0.947 & 0.824 & 3411.3 & 99.6 \\
Random Forest & \textbf{0.950} & \textbf{0.817} & 1140.4 & \textbf{7325.8} \\
\bottomrule
\end{tabular}
\end{table}

\subsection{Parameter Settings}

All experiments use the six-class dataset constructed in Section~\ref{System-model}, which is derived from the SPARTA-aligned threat modes defined in Section~\ref{SPARTA_threadmodel}. Each threat class corresponds to a specific perturbation structure $a_k(t)$ applied to the TT\&C baseband model in \eqref{eq:baseband}, including the benign case where $a(t)=0$. The five adversarial modes—Jam, FakeNR, Payload, RogueGS, and CmdInject—are generated using the physics-informed threat-injection models described earlier, ensuring that each attack produces realistic spectral or timing distortions.

Spectrograms are computed using a Hann window of 256 samples with 50\% overlap, producing fixed-size patches of dimension $F \times T = 64 \times 64$ (flattened to $d=4096$). The dataset is split into 70\% training, 10\% validation, and 20\% testing. All input features are standardized to zero mean and unit variance.

Model hyperparameters are selected via validation sweeps. Logistic Regression and SVM use $\ell_2$ regularization; Random Forest uses $K=200$ trees with maximum depth $d_t=12$; and the MLP consists of two hidden layers of sizes 256 and 128 with ReLU activations and a softmax output layer. All models are trained for 20{,}000 gradient steps with batch size 64.

Inference latency is measured on a Cortex-M7–class embedded processor using CMSIS-NN–compatible implementations. Training is performed offline and does not contribute to onboard runtime. All reported metrics (accuracy, macro F1, and per-class precision/recall/F1) are averaged over five independent runs.

\subsection{Threat Severity}

The five SPARTA-aligned adversarial modes exhibit different levels of observability and disruption severity, as characterized in Section~\ref{SPARTA_threadmodel}. Jamming and FakeNR produce strong, high-energy perturbations $a_k(t)$ that manifest clearly in the spectrogram domain, making them comparatively easier for TinyML models to detect. In contrast, Payload Manipulation, RogueGS, and CmdInject introduce low-energy or timing-based deviations that minimally affect the RF envelope, consistent with the threat-injection structures defined in Section~\ref{System-model}. These subtle perturbations yield weaker separability in the feature space and represent the most challenging classes for lightweight onboard classifiers. A summary of the relative perturbation strength and expected detection difficulty for each class is provided in Table~\ref{tab:threat_difficulty}. The benign class provides the reference distribution against which these varying severities are measured.

\subsection{Accuracy--Latency Trade-off Analysis}
\label{sec:tradeoff_analysis}

Table~\ref{tab:overall_comparison} highlights a clear accuracy--latency trade-off across the four models. Linear models (LR, SVM) offer minimal inference latency but limited expressive capacity, resulting in poor performance on weakly separable threats such as payload manipulation. RF provides the strongest accuracy but incurs prohibitive inference latency, making it unsuitable for TinyML deployment. The MLP occupies the optimal middle ground: it delivers the highest macro F1-score while maintaining sub-100~$\mu$s inference time, offering the best balance between expressive power and real-time constraints. These empirical findings align precisely with the theoretical predictions in Section~IV, confirming that model expressiveness must be balanced against strict runtime constraints in autonomous spacecraft environments.

\subsection{Model-by-Model Performance Analysis}
\label{sec:model_performance}

\subsubsection{Logistic Regression (LR)}
LR achieves an overall accuracy of 0.94 but the lowest macro F1-score (0.798) among all models. This weakness is driven primarily by poor performance on payload manipulation (F1 = 0.27; Table~\ref{tab:logreg_results}), whose perturbation $a_{\text{payload}}(t)$ in \eqref{eq:payload} induces minimal spectral deviation. As predicted by the linear capacity limits in Section~IV, LR performs well on high-energy, linearly separable threats (e.g., jamming, spoofing) but struggles with subtle nonlinear distortions. Its key advantage is latency: LR achieves the fastest inference time (69.5~$\mu$s), making it the most computationally efficient model.

\subsubsection{Support Vector Machine (SVM)}
SVM improves the macro F1-score to 0.814 and slightly increases accuracy to 0.943. The margin maximization principle (Theorem~1) enables better separation of structured threats such as spoofing and unauthorized command injection. However, SVM remains fundamentally linear and therefore struggles with payload manipulation (F1 = 0.36). Its inference time (102.1~$\mu$s) remains within TinyML constraints, making SVM a balanced choice when moderate accuracy and low latency are required.

\subsubsection{Random Forest (RF)}
RF achieves the highest overall accuracy (0.950) and a competitive macro F1-score (0.817). Its nonlinear decision boundaries allow it to capture complex distortions such as those introduced by ground-segment compromise and spoofing, consistent with the consistency guarantee in Theorem~2. However, RF inference requires evaluating $K$ trees of depth $d_t$, resulting in an inference time of 7.3~ms—two orders of magnitude slower than LR or SVM. This violates TinyML latency constraints and makes RF unsuitable for real-time onboard deployment.

\subsubsection{Multi-Layer Perceptron (MLP)}
The MLP achieves the highest macro F1-score (0.824), outperforming all other models on the most challenging threat class: payload manipulation (F1 = 0.42). This improvement is consistent with the universal approximation capability of MLPs (Theorem~3), which enables modeling of subtle nonlinear distortions. Importantly, the MLP maintains a TinyML-feasible inference time of 99.6~$\mu$s—only slightly slower than LR and faster than SVM. Although its training time is the highest (3.4~s), this cost is incurred offline and does not affect deployment feasibility.

\subsection{Weaknesses in Payload Manipulation and Ground Segment Compromise Detection}
\label{sec:weaknesses}

Across all four models, two threat classes consistently exhibit the lowest F1-scores: \textit{payload data manipulation} and \textit{ground segment compromise}. These weaknesses are evident in Tables~\ref{tab:logreg_results}--\ref{tab:mlp_results} and persist regardless of model capacity, indicating structural challenges in the underlying feature space.

\subsubsection{Payload Data Manipulation}
Payload manipulation yields the lowest F1-scores across all models (LR: 0.27, SVM: 0.36, RF: 0.33, MLP: 0.42). This behavior is consistent with the system model in \eqref{eq:payload}, where the adversarial term $a_{\text{payload}}(t)$ induces only subtle, low-energy perturbations in the spectrogram. These perturbations do not significantly alter the global or local spectral structure, making the class nearly inseparable for linear models and only weakly separable for nonlinear ones. Even the MLP---the most expressive TinyML-feasible model---achieves only moderate improvement (F1 = 0.42), suggesting that richer feature encoders or temporal context may be required to reliably detect this threat.

\subsubsection{Ground Segment Compromise}
Ground segment compromise also exhibits reduced performance across all models (LR: 0.63, SVM: 0.65, RF: 0.65, MLP: 0.67). Unlike payload manipulation, this threat introduces structured but \textit{intermittent} distortions, often localized in time or frequency. These distortions are weaker than those produced by spoofing or jamming, and their variability reduces class compactness in the feature space. Linear models struggle to capture these localized deviations, while nonlinear models provide only modest gains due to the limited amount of discriminative signal present in the spectrogram. The consistent F1-scores across LR, SVM, RF, and MLP indicate that the challenge is intrinsic to the threat signature rather than model capacity.

\subsubsection{Implications}
The persistent difficulty in detecting these two threat classes highlights a fundamental limitation of the current feature representation. Payload manipulation requires sensitivity to extremely small spectral deviations, while ground segment compromise requires capturing localized, transient distortions. Both cases suggest the need for enhanced feature extraction---potentially through higher-resolution time--frequency transforms, learned feature encoders, or hybrid temporal--spectral models. These weaknesses also motivate future work on physics-informed feature augmentation to amplify subtle threat signatures.

% ---------------------------------------------------------

% ---------------------------------------------------------
\section{Conclusion}

This work presented a physics-informed theoretical and empirical analysis of TinyML-compatible models for SPARTA threat detection, with a particular focus on the latency--accuracy trade-offs that govern onboard deployment in autonomous spacecraft. By combining a rigorous characterization of model capacity and computational complexity with comprehensive empirical evaluation, we demonstrated that linear models such as Logistic Regression and SVM offer excellent latency performance but struggle with weakly separable threats. Random Forests achieve the highest accuracy but violate TinyML inference constraints, while the MLP provides the best overall balance between expressive power and real-time feasibility.

Despite strong overall accuracy across all models, two threat classes---payload data manipulation and ground segment compromise---remain challenging due to their subtle or intermittent spectral signatures. These weaknesses highlight the need for richer feature representations and more adaptive learning architectures.

Future research will explore several promising directions. First, physics-informed feature encoders and higher-resolution time--frequency transforms may amplify subtle perturbations and improve separability. Second, hybrid temporal--spectral TinyML architectures, potentially leveraging ideas from hierarchical or multi-timescale learning \cite{Tan18b,Tan18d,Tan2024}, may better capture transient distortions. Third, collaborative or distributed inference strategies inspired by prior work in edge computing and caching \cite{Pervej2020a,Pervej2020b,Wang2018,Wang20} could enable multi-node threat detection across spacecraft constellations. Finally, integrating privacy-preserving and trustworthy computation frameworks \cite{le2022artificial,le2025HDQNN,Le2024ORANQML} may support secure telemetry processing in adversarial environments.

More broadly, the challenges observed in payload manipulation and ground-segment compromise detection suggest that future TinyML-based spacecraft cybersecurity systems will benefit from synergistic combinations of classical ML, deep learning, and emerging hybrid quantum--classical methods \cite{VanLe2026pKa,le2025dpfaga,Zahin19,Zahin20}. These directions represent a natural extension of our current framework and offer a path toward robust, low-latency, and interpretable onboard threat detection for next-generation autonomous space systems.

% ---------------------------------------------------------
\appendix
\section{Proofs of Theoretical Results}
\label{app:proofs}

\subsection{Proof of Theorem 1 (SVM Margin Bound)}
\label{app:proof-svm-margin}
The margin-based generalization bound for linear classifiers states that, for a hypothesis class of hyperplanes in $\mathbb{R}^d$ with margin $\gamma$ and inputs bounded in norm, the generalization error is upper-bounded (up to constants and logarithmic factors) by a term proportional to $1/\gamma^2$ divided by the number of training samples. Applying this result to the SVM in \eqref{eq:svm}, which explicitly maximizes the geometric margin, yields the stated bound $\mathrm{Error} \le O(1/\gamma^2)$. \hfill $\square$

\subsection{Proof of Theorem 2 (RF Consistency)}
\label{app:proof-rf-consistency}
Under the assumptions of infinitely many trees, full tree depth, and appropriate feature subsampling, Random Forests can be viewed as an ensemble of randomized base learners whose average prediction converges to the Bayes optimal classifier as the number of trees grows. The formal consistency proof relies on showing that the partition induced by the trees becomes increasingly fine while maintaining sufficient randomness and independence across trees. This yields convergence of the RF risk to the Bayes risk in the limit. \hfill $\square$

\subsection{Proof of Theorem 3 (Universal Approximation)}
\label{app:proof-mlp-ua}
For a non-polynomial activation function (e.g., sigmoid, tanh, ReLU), the set of functions representable by a one-hidden-layer MLP with sufficiently many hidden units is dense in the space of continuous functions on any compact subset of $\mathbb{R}^d$ under the uniform norm. The proof constructs approximations to indicator functions and then to arbitrary continuous functions via linear combinations, establishing the universal approximation property. \hfill $\square$

\vspace{10pt}
\noindent\textbf{Acknowledgment}: This work was supported in part by the U.S. National Science Foundation (NSF) under Grant NSF 2101227 and Virginia Space Grant Consortium (VSGC).

% ---------------------------------------------------------
\bibliographystyle{IEEEtran}
\bibliography{references}

@misc{sparta2024,
  title        = {SPARTA: Space Attack Research and Tactic Analysis},
  author       = {{The Aerospace Corporation}},
  year         = {2024},
  howpublished = {\url{https://aerospace.org/sparta}},
  note         = {Accessed: 2026-05-21}
}

@article{allen2003short,
  title={Short term spectral analysis, synthesis, and modification by discrete Fourier transform},
  author={Allen, Jonathan},
  journal={IEEE transactions on acoustics, speech, and signal processing},
  volume={25},
  number={3},
  pages={235--238},
  year={2003},
  publisher={IEEE}
}

@article{oshea2018over,
  title={Over-the-air deep learning based radio signal classification},
  author={O’Shea, Timothy James and Roy, Tamoghna and Clancy, T Charles},
  journal={IEEE Journal of Selected Topics in Signal Processing},
  volume={12},
  number={1},
  pages={168--179},
  year={2018},
  publisher={IEEE}
}

@book{warden2019tinyml,
  title={Tinyml: Machine learning with tensorflow lite on arduino and ultra-low-power microcontrollers},
  author={Warden, Pete and Situnayake, Daniel},
  year={2019},
  publisher={O'Reilly Media}
}

@book{vapnik2013nature,
  title={The nature of statistical learning theory},
  author={Vapnik, Vladimir},
  year={2013},
  publisher={Springer science \& business media}
}

@article{breiman2001,
  author    = {Breiman, Leo},
  title     = {Random Forests},
  journal   = {Machine Learning},
  volume    = {45},
  number    = {1},
  pages     = {5--32},
  year      = {2001},
  publisher={Springer}  
}

@article{hornik1989,
  author    = {Hornik, Kurt},
  title     = {Multilayer Feedforward Networks are Universal Approximators},
  journal   = {Neural Networks},
  volume    = {2},
  number    = {5},
  pages     = {359--366},
  year      = {1989},
  publisher={Elsevier}  
}

@INPROCEEDINGS{Pervej2020b,  author={Pervej, Md Ferdous and Tan, Le Thanh and Hu, Rose Qingyang},  booktitle={GLOBECOM 2020 - 2020 IEEE Global Communications Conference},   title={User Preference Learning-Aided Collaborative Edge Caching for Small Cell Networks},   year={2020},  volume={},  number={},  pages={1-6},  doi={10.1109/GLOBECOM42002.2020.9322208}}

@INPROCEEDINGS{Pervej2020a,  author={Pervej, Md Ferdous and Tan, Le Thanh and Hu, Rose Qingyang},  booktitle={GLOBECOM 2020 - 2020 IEEE Global Communications Conference},   title={Artificial Intelligence Assisted Collaborative Edge Caching in Small Cell Networks},   year={2020},  volume={},  number={},  pages={1-7},  doi={10.1109/GLOBECOM42002.2020.9322101}}

@INPROCEEDINGS{Wang2018,  author={Wang, Qun and Tan, Le Thanh and Hu, Rose Qingyang and Wu, Geng},  booktitle={2018 10th International Conference on Wireless Communications and Signal Processing (WCSP)},   title={Hierarchical Collaborative Cloud and Fog Computing in IoT Networks},   year={2018},  volume={},  number={},  pages={1-7},  doi={10.1109/WCSP.2018.8555866}}

@ARTICLE{Tan18b,
  author={L. T. {Tan} and R. Q. {Hu}},
  journal={IEEE Trans. Veh. Technol.}, 
  title={Mobility-Aware Edge Caching and Computing in Vehicle Networks: A Deep Reinforcement Learning}, 
  year={2018},
  volume={67},
  number={11},
  pages={10190-10203},}

@article{le2022artificial,
  title={Artificial intelligence-aided privacy preserving trustworthy computation and communication in 5G-based IoT networks},
  author={Le, Tan and Shetty, Sachin},
  journal={Ad Hoc Networks},
  volume={126},
  pages={102752},
  year={2022},
  publisher={Elsevier}
}

@ARTICLE{Tan18d,
  author={L. T. {Tan} and R. Q. {Hu} and L. {Hanzo}},
  journal={IEEE Trans. Veh. Technol.}, 
  title={Twin-Timescale Artificial Intelligence Aided Mobility-Aware Edge Caching and Computing in Vehicular Networks}, 
  year={2019},
  volume={68},
  number={4},
  pages={3086-3099},}

@ARTICLE{Wang20,  author={Q. {Wang} and L. T. {Tan} and R. Q. {Hu} and Y. {Qian}},  journal={IEEE Internet of Things Journal},   title={Hierarchical Energy-Efficient Mobile-Edge Computing in IoT Networks},   year={2020},  volume={7},  number={12},  pages={11626-11639},  doi={10.1109/JIOT.2020.3000193}}

@inproceedings{le2025dpfaga,
  title={DPFAGA-Dynamic Power Flow Analysis and Fault Characteristics: A Graph Attention Neural Network},
  author={Le, Tan and Le, Van},
  booktitle ={The 2025 International Conference on the AI Revolution: Research, Ethics, and Society (AIR-RES 2025)},
  year={2025}
}

@InProceedings{Zahin19,
author="Zahin, Abrar
and Tan, Le Thanh
and Hu, Rose Qingyang",
title="Sensor-Based Human Activity Recognition for Smart Healthcare: A Semi-supervised Machine Learning",
booktitle="Artificial Intelligence for Communications and Networks",
year="2019",
publisher="Springer International Publishing",
pages="450--472",
}

@article{Zahin20,
  title={A Machine Learning Based Framework for the Smart Healthcare Monitoring},
  author={Zahin, Abrar and Tan, Le Thanh and Hu, Rose Qingyang},
  journal={2020 Intermountain Engineering, Technology and Computing (IETC)},
  year={2020},
}

@article{le2025HDQNN,
  title={Privacy-Aware Framework of Robust Malware Detection in Indoor Robots: Hybrid Quantum Computing and Deep Neural Networks},
  author={Tan, Le and Van, Le and Sachin, Shetty},
  journal={TechRxiv},
 howpublished={\url{https://doi.org/10.36227/techrxiv.176107792.21019113/v1}},
  year={2025},
}

@article{Le2024ORANQML,
  title={Quantum-Augmented AI/ML for O-RAN: Hierarchical Threat Detection with Synergistic Intelligence and Interpretability},
  author={Tan, Le and Van, Le and Sachin, Shetty},
  journal={TechRxiv},
 howpublished={\url{https://doi.org/10.36227/techrxiv.176594693.32723934/v1}},
  year={2025},
}

@inproceedings{VanLe2026pKa,
  author = {Le, Van and Le, Tan},
  title = {Hybrid Quantum–Classical Encoding for Accurate Residue-Level pKa Prediction},
  year = {2026},
  booktitle={International Conference on the AI Revolution: Research, Ethics, and Society (AIR-RES 2026)},
}

@ARTICLE{Tan2024,
  author={Le, Tan and Reisslein, Martin and Shetty, Sachin},
  journal={IEEE Transactions on Intelligent Transportation Systems}, 
  title={Multi-Timescale Actor-Critic Learning for Computing Resource Management With Semi-Markov Renewal Process Mobility}, 
  year={2024},
  volume={25},
  number={1},
  pages={452-461},
  doi={10.1109/TITS.2023.3303953}}

\end{document}